\documentclass[superscriptaddress,
  preprint,
  showpacs,
  pra]{revtex4}
\begin{document}

\title{Doppler effect in Schwarzschild geometry}

\author{A.\ Radosz}
\affiliation{Institute of Physics, Wroc\l aw University of Technology,
  Wybrze\.ze Wyspia\'nskiego 27, 50-370 Wroc\l aw, Poland}
\author{A.\ T.\ Augousti}
\affiliation{Faculty of Science, Kingston University,
  Kingston, Surrey KT1 2EE UK}
\author{K.\ Ostasiewicz}
\affiliation{Institute of Physics, Wroc\l aw University of Technology,
  Wybrze\.ze Wyspia\'nskiego 27, 50-370 Wroc\l aw, Poland}

\begin{abstract}
The Doppler shift considered in general relativity involves mixed
contributions of distinct, gravitational and kinematical origins and for most
metrics or trajectories it takes a complex form. The expression for the
Doppler shift may simplify due to particular symmetries. In Schwarzschild
spacetime it factorizes in the case of radial fall for an observer and radial
null geodesic. The resulting expression is composed of factors that can be
identified with contributions arising from classical, special relativistic and
general relativistic origins. This result turns out to be more general: it
holds for the whole class of observers travelling parallel to the spatial path
of null geodesics when receiving the signal. It also holds for a particular
type of in-fall in the case of a Kerr metric.
\end{abstract}


\maketitle

In the presence of a gravitational field, the Doppler shift may be regarded as
composed of three distinct contributions from different origins: the classical
effect and its relativistic modification (``time dilation''), both of
kinematic character, and a gravitational red/blue shift. The combined effect
in the case of general spacetime geometries, is in general a complex
coupling of these contributions. In this paper we confine ourselves to a
discussion of the Doppler shift in a Schwarzschild spacetime. We demonstrate
that its coupled form factorizes in the particular case of free radial fall.
Such a decoupling is proven to be more general: it holds for an arbitrary
radial in-fall and in fact for the whole class of trajectories tangential to a
null geodesic in this geometry. It also holds for a fall along an axis of
symmetry in the case of Kerr geometry.

Let us begin the discussion from a definition of a Schwarzschild spacetime,
where the line element expressed in Schwarzschild coordinates takes the form:
\begin{eqnarray}
  ds^2&=&\left(1-\frac{r_S}{r}\right)dt^2-
         \left(1-\frac{r_S}{r}\right)^{-1}dr^2
\nonumber\\
       &&{}-r^2d\theta^2-r^2\sin^2\theta\,d\varphi^2
\nonumber\\
 &\equiv&g_{tt}dt^2+g_{rr}dr^2+g_{\theta\theta}d\theta^2
         +g_{\varphi\varphi}d\varphi^2.
\end{eqnarray}
There are two Killing vectors in this case, $\tilde{\eta}$ and $\tilde{\xi}$,
\begin{equation}
  \eta^\alpha=\delta^\alpha_t, \quad \xi^\alpha=\delta^\alpha_\varphi,
\label{killing}
\end{equation}
corresponding to energy and angular momentum conservation, respectively.
A {\it massless particle\/} (e.g.\ photon) follows planar null geodesics,
taken here as equatorial ($\theta=\pi/2$):
\begin{equation}
  \kappa^\alpha\kappa_\alpha=0,
\end{equation}
where a tangential vector (wave vector) is denoted as $\tilde{\kappa}$. Due to
conservation laws, (\ref{killing}), $\tilde{\kappa}'s$ components are,
\begin{equation}
  \kappa^t=\frac{\epsilon}{g_{tt}}, \quad
  \kappa^r=\pm\sqrt{\epsilon^2-g_{tt}\frac{\lambda^2}{r^2}}, \quad
  \kappa^\varphi=\frac{\lambda}{r^2},
\end{equation}
where $\epsilon$ and $\lambda$ are conserved quantities. An arbitrary
observer, hereafter termed A, follows a world line, characterized by a
time-like velocity vector $\tilde{U}$:
\begin{equation}
  U^\alpha U_\alpha =1.
\end{equation}
A static observer, denoted hereafter as S, provides the absolute rest system,
and his velocity is a normalized Killing vector $\tilde{\eta}$ (see also
\cite{cite1}):
\begin{equation}
  \tilde{N}=(\eta^\alpha\eta_\alpha)^{-1/2}\tilde{\eta}, \quad
  \eta^\alpha\eta_\alpha=g_{tt}.
\end{equation}
Observer A, travelling in the equatorial plane, receiving a light signal
measures its frequency as the time-component of the $\tilde{\kappa}$ vector:
\begin{equation}
  \omega^*=U^\alpha\kappa_\alpha=
  g_{tt}U^t\kappa^t\left(1+\frac{U^i\kappa_i}{U^t\kappa_t}\right),
\label{lightsignal}
\end{equation}
where the notation,
\begin{equation}
  U^t\kappa_t\equiv g_{tt}U^t\kappa^t, \quad
  U^i\kappa_i\equiv g_{rr}U^r\kappa^r+
                    g_{\varphi\varphi}U^\varphi\kappa^\varphi
\end{equation}
has been applied. In particular, observer S measures the light signal of
frequency:
\begin{equation}
  \overline{\omega}=N^\alpha\kappa_\alpha=\sqrt{g_{tt}}\kappa^t,
\label{freqS}
\end{equation}
which corresponds to the gravitational blue (or red -- see below) shift. The
Doppler shift arises when the frequency of the light signal, as measured by A
or S, Eqs. (\ref{lightsignal}), (\ref{freqS}), is compared with its emission
frequency. We consider here the case of a static source of the light signal.
If the light was emitted at infinity, by an inertial observer, its initial
frequency is $\omega_\infty\equiv\epsilon$; when it is emitted by a static
source located at radial coordinate $r_1$, its frequency $\omega_1$ is related
to $\omega_\infty$ by
\begin{equation}
  \omega_1=\frac{\omega_\infty}{\sqrt{g_{tt}(r_1)}}.
\label{relates}
\end{equation}
The velocity $\tilde{V}$ of observer A with respect to the static observer S,
(see e.g.\ \cite{cite1}), is given by
\begin{equation}
  \tilde{V}=\tilde{U}-(U^\alpha N_\alpha)\tilde{N}=(0,U^i)
\label{eq11}
\end{equation}
and is related to the {\it speed}, $v$, (measured by S) as follows:
\begin{equation}
  U^iU_i=-\frac{v^2}{1-v^2}\equiv -(\gamma v)^2.
\label{relate1}
\end{equation}
Observations (\ref{relates}) and (\ref{relate1}) lead to the final form for
the frequency of the light received at $r$ by the observer A, traveling with
speed $v$ (see also \cite{cite2}):
\begin{equation}
  \omega^*=\gamma\frac{\omega_\infty}{\sqrt{g_{tt}(r)}}
  \left( 1+\frac{U^i\kappa_i}{U^t\kappa_t}\right).
\label{doppler1}
\end{equation}
This formula provides the Doppler shift, for the signal emitted both at
infinity and at finite radial coordinate $r_1$. Though expression
(\ref{doppler1}) has the form of a product, the bracketed factor mixes up
velocity components and wave vector components, thus resulting in an
entangled form. Under what circumstances would factorization of this
expression into classical, special relativistic and general relativistic
contributions occur? The answer for this is the following. Noticing the
relation (see (\ref{relate1})):
\begin{equation}
  v^2=-\frac{U^iU_i}{U^tU_t}
\end{equation}
one can identify the case when frequency (\ref{doppler1}) simplifies to the
expected form. If the spatial components of velocity vector, $U^i$, and the
components of wave vector, $\kappa^i$, are {\it parallel\/} (or
{\it anti-parallel\/}), i.e.
\begin{equation}
  (U^i\kappa_i)^2=(U^iU_i)\cdot (\kappa^i\kappa_i),
\label{parallel}
\end{equation}
then the bracket in Eq. (\ref{doppler1}) simplifies:
\begin{equation}
  \left(1+\frac{U^i\kappa_i}{U^t\kappa_t}\right)=1\mp v.
\end{equation}
In this case, the Doppler shift for the light signal emitted at infinity
takes a factorized form:
\begin{equation}
  \frac{\omega^*}{\omega_\infty}=\frac{1}{\sqrt{g_{tt}(r)}}
  \frac{\sqrt{1-v^2}}{1\pm v},
\label{shift1}
\end{equation}
where kinematical (classical and relativistic) and gravitational contributions
are clearly identified. It is modified to a more symmetric expression for the
case of a light signal emitted at $r_1$:
\begin{equation}
  \frac{\omega^*}{\omega_1}=\frac{\sqrt{g_{tt}(r_1)}}{\sqrt{g_{tt}(r)}}
  \frac{\sqrt{1-v^2}}{1\pm v}.
\label{shift2}
\end{equation}

Let us briefly consider three particular cases of null geodesics.

Consider first the case of radial fall. An in-falling observer, who
starts free fall at infinity, and receives a light signal emitted from his
starting point, finds particular cancellation of the kinematical ({\it time
dilation\/}) and {\it gravitational\/} (blue shift) contributions (see also
\cite{eur}):
\begin{equation}
  \frac{\omega^*}{\omega_\infty}=\frac{1}{1+v}.
\label{radial1}
\end{equation}
This observation has an obvious generalization for a free fall from a finite
distance, $r_1$. As the light signal is emitted from $r_1$ the same
cancellation occurs:
\begin{equation}
  \frac{\omega^*}{\omega_1}=\frac{\sqrt{g_{tt}(r_1)}}{\sqrt{g_{tt}(r)}}
  \frac{\sqrt{1-v^2}}{1+v}\equiv\frac{1}{1+v}.
\label{radial2}
\end{equation}
The meaning of this effect is the following. Massive and massless objects,
following radial geodesics, change their energies in the same way: the ratio
of photon energies at points $r_1$ and $r$ is the same as the massive object
energy ratio taken at these points. If a body is released from an initial
rest, then the classical form of Doppler shift (\ref{radial1}),
(\ref{radial2}) is obtained; initial non zero velocity $v_0$ results in a
residual time dilation factor, $\sqrt{1-v_0^2}$, in (\ref{radial1}),
(\ref{radial2}).

Another special example of a light geodesic is the ``photon sphere'' (see also
\cite{cite1}, \cite{cite3}). The light orbits freely on a circle of radius
$r=(3/2)r_S$ and when an observer A is constrained to move along that circle
(for which the acceleration is independent of the speed),
\begin{equation}
  \tilde{U}=B\tilde{\eta}+D\tilde{\xi}.
\end{equation}
Then, as condition (\ref{parallel}) is satisfied, during uniform circulation,
$B=\mathord{\rm const}$, $D=\mathord{\rm const}$, the factorized Doppler shift
(\ref{shift2}) simplifies to the residual form:
\begin{equation}
  \frac{\omega^*}{\omega_0}=\frac{\sqrt{1-v^2}}{1\pm v},
\end{equation}
where the constant velocity is equal to
\begin{equation}
  v^2=\frac{D^2}{B^2}\frac{r^2}{g_{tt}}.
\end{equation}
The absence of the gravitational contribution is obvious:
\begin{equation}
  \frac{\sqrt{g_{tt}(r_0)}}{\sqrt{g_{tt}(r)}}=1.
\end{equation}

The third example leads to peculiar observation. Let us consider radial photon geodesic,
$\kappa^i=(\kappa^r,0,0)$, and observer A, who revolves around the centre, $U^i=(0,0,U^\varphi )$.
 Then the ``scalar product'' , $U^i \kappa_i =0$, vanishes and one finds, according Eq. (\ref{eq11})
\begin{equation}
\frac{\omega^\star}{\omega_0}=\frac{1}{\sqrt{g_{tt}(r)}}\frac{1}{\sqrt{1-v^2}}.
\end{equation}
This result appears to be counter intuitive: observer A finds gravitational
blueshift and also blue shift of kinematical origin (!), instead of ordinary
time-dilation-redshift.

The final conclusion is the following. In Schwarzschild spacetime, for a
static source of the light signal, the Doppler shift may be decomposed into
the form of three factors of distinct origins. It occurs in the case for an
observer travelling tangentially to the the null geodesics when receiving the
signal. In a particular realization, the situation of free fall of a receiving
observer, gravitational blue shift and time-dilation red shift cancel each
other out, in fact reflecting energy conservation. Let us mention that the
effect (\ref{shift1}), (\ref{shift2}) also holds in the case of Kerr
spacetime, for a particular type of null geodesic, namely for the axis of
symmetry and motion of an observer along this axis. It should be underlined
that the special property,
\begin{equation}
  g_{tt}\cdot g_{rr}=-1,
\end{equation}
is common for Schwarzschild geometry and for that particular motion along the
axis of symmetry in Kerr geometry.

\end{document}